\documentclass[aps,pre,reprint,superscriptaddress,longbibliography]{revtex4-2}

\usepackage{graphicx} 
\usepackage{amsmath}
\usepackage{comment}
\usepackage{physics}
\usepackage[dvipsnames]{xcolor}
\usepackage{amsfonts}
\usepackage[normalem]{ulem} 

\begin{document}
	\newcommand{\K}{{\mathrm K}}
	\newcommand{\ud}{{\mathrm d}}
	\newcommand{\umod}{\mathrm{mod}}


\title{Quantum reinforcement learning in the presence of thermal dissipation}


\author{Mar\'{i}a Laura Olivera-Atencio}
\affiliation{F\'{\i}sica Te\'orica,Universidad de Sevilla, Apartado de Correos 1065, Sevilla 41080, Spain}
\author{ Lucas Lamata}
\affiliation{Departamento de F\'isica At\'omica, Molecular y Nuclear, Universidad de Sevilla, 41080 Sevilla, Spain}
\affiliation{Instituto Carlos I de F\'isica Te\'orica y Computacional, 18071 Granada, Spain}
\author{ Manuel Morillo}
\affiliation{F\'{\i}sica Te\'orica,Universidad de Sevilla, Apartado de Correos 1065, Sevilla 41080, Spain}
\author{Jes\'us Casado-Pascual}
\email[]{jcasado@us.es}
\affiliation{F\'{\i}sica Te\'orica,Universidad de Sevilla, Apartado de Correos 1065, Sevilla 41080, Spain}


\date{\today}

\begin{abstract}
A study of the effect of thermal dissipation on quantum reinforcement learning is performed. For this purpose, a nondissipative quantum reinforcement learning protocol is adapted to the presence of thermal dissipation. Analytical calculations as well as numerical simulations are carried out obtaining evidence that dissipation do not significantly degrade the performance of the quantum reinforcement learning protocol for sufficiently low temperatures, being in some cases even beneficial. Quantum reinforcement learning under realistic experimental conditions of thermal dissipation opens an avenue for the realization of quantum agents able to interact with a changing environment, and adapt to it, with plausible many applications inside quantum technologies and machine learning. 
\end{abstract}


\maketitle


\section{Introduction}
Quantum machine learning~\cite{BiamonteNature,LamataReview} aims at employing quantum technologies for achieving machine learning tasks more efficiently. 
 A diversity of quantum algorithms for machine learning have been proposed, 
including solvers for linear systems of equations~\cite{PhysRevLett.103.150502}, quantum principal component analysis~\cite{NatPhysQPCA}, quantum support vector machines~\cite{PRLettQSVM}, quantum annealers~\cite{ReviewQAnnealing}, quantum variational eigensolvers~\cite{ReviewQEigensolver}, quantum reinforcement learning~\cite{Dong2008,Paparo2014,Dunjko2016,Bukov2018,Bukov_Day2018,Fosel2018,Liu2022,Albarran_Arriagada_2020}, quantum generative adversarial networks~\cite{QGAN}, as well as quantum kernels~\cite{QKernels}.  Experimental implementations showing, in some instances, speedups with respect to classical or other kinds of quantum algorithms have been carried out already at the noisy intermediate scale quantum (NISQ) devices~\cite{QGAN,QKernels,QReinfHefei,QReinfVienna,QMemristVienna,QAdvGoogle}.  Many of these works study the advantages of quantum technologies over classical machine learning. However, the effect of thermal dissipation on quantum machine learning has not yet been analyzed in depth, although performing a real quantum experiment always involves working at nonzero temperature and, therefore, a certain amount of such dissipation. Some preliminary results on this issue can be found in Refs.~\cite{Nguyen2020,Lu2022}. 
	
 In the field of quantum machine learning, as in the classical case, algorithms can be classified into three main groups: supervised~\cite{Innocenti_2020,Youssry_2020,Shrapnel_2018,Luchnikov_2020,Melnikov_2020,Melnikov_2019,Kalantre_2019,Torlai_2019,Torlai_2018}, unsupervised~\cite{Liu_2020,Agresti_2019}, and reinforcement quantum algorithms~\cite{Dong2008,Paparo2014,Dunjko2016,Bukov2018,Bukov_Day2018,Fosel2018,Liu2022,Albarran_Arriagada_2020}. In this paper, we will focus on the latter group and, more specifically, on the effects of thermal dissipation on this type of algorithms. Reinforcement learning algorithms consist on successive interactions  between a known agent and an unknown environment alternated with a reward function that improves a specific task performed by such agent~\cite{Sutton2018,DMDP}. The goal is to learn from the environment. The information extraction from the environment as well as the channel used to communicate the information to the agent and the action of the agent, are established by the policy. The reward function defines the criterion used to punish or reward certain actions of the agent in order to improve its performance. Each reward reinforces the current strategy while the punishment forces an adaptation of the strategy.

 To study the effect of dissipation on quantum reinforcement learning, in this paper we adapt the nondissipative algorithm proposed in Ref.~\cite{Albarran_Arriagada_2020} to the presence of thermal dissipation. For this purpose, the environment will be considered to be an unknown quantum system whose dynamics, besides including a unitary part associated with a given Hamiltonian as in~\cite{Albarran_Arriagada_2020}, also possesses a certain amount of thermal dissipation. The goal is to extract information or learn from the environment to obtain near-optimal knowledge of the eigenstates of the Hamiltonian. The agent corresponds to a known and manipulable quantum state that must conveniently adjust to the dynamics of the environment in order to approach the unknown eigenstate. It should be noted that the aim of this work is not to analyze the scalability or speedup of this type of protocols, which has been discussed previously~\cite{QReinfHefei,Albarran_Arriagada_2020}, but to study the effect of thermal dissipation on it. For this reason, and for the sake of clarity, we will focus on the simplest case of a two-level quantum system.

The structure of the remainder of this paper is as follows. In Sec.~\ref{protocol},  the nondissipative protocol proposed in Ref.~\cite{Albarran_Arriagada_2020} is adapted to the presence of thermal dissipation. Special emphasis is given to the peculiarities associated with the presence of such dissipation. In Sec.~\ref{numerics}, we discuss the numerical implementation of the protocol and illustrate our results with numerical simulations. Finally, in Sec.~\ref{Conclusions}, we present conclusions for the main findings of our work.

\section{Protocol description}

\label{protocol}

In the  nondissipative protocol proposed in Ref.~\cite{Albarran_Arriagada_2020}, the agent \texttt{A}  is considered to be a known manipulable quantum system described by a state vector $\ket{\phi}$. The environment \texttt{E} is modeled as a ``black box'' that interacts with \texttt{A} for a time $\tau$. The effect of this interaction on \texttt{A} is characterized by a unitary transformation $U \equiv e^{-i \tau H/\hbar}$ applied to the state vector $\ket{\phi}$, where $H$ is an unknown interaction Hamiltonian whose eigenvectors are to be computed. For simplicity, we will henceforth restrict the analysis to the case of a single qubit  with state basis vectors $\{\ket{0}, \ket{1}\}$ and unknown interaction Hamiltonian
\begin{equation}
	\label{Hamiltonian}
	H=\frac{\hbar \omega}{2}\left(\ketbra{+}{+}-\ketbra{-}{-}\right),
\end{equation}
where $\omega$ is a positive constant with dimensions of frequency and $\{\ket{+}, \ket{-}\}$ are the eigenvectors to be computed.

Suppose now that, in addition to interacting with \texttt{E}, \texttt{A} is also in contact with a thermal bath \texttt{B} at a finite temperature $T$. 
We will also assume that the combined action of \texttt{E} and \texttt{B} on \texttt{A} is described by a Lindblad master equation of the form~\cite{BreuerPetruccione2003}
\begin{equation}
	\label{Lindblad_eq}
	\dot{\rho}_t= -\frac{i}{\hbar} [H,\rho_t]+\sum_{j=\pm} \Gamma_j\left(\tilde{\sigma}_j^{\dagger}\rho_t\tilde{\sigma}_j-\frac{1}{2}\{\tilde{\sigma}_j\tilde{\sigma}_j^{\dagger},\rho_t\}\right),
\end{equation}
where $\rho_t$ is the density operator representing the state of \texttt{A} at time $t$, $\smash{\tilde{\sigma}_-=\ketbra{-}{+}=\tilde{\sigma}_+^{\dagger}}$ is a Lindblad operator that induces dissipative decay from the excited state $\ket{+}$ to the ground state $\ket{-}$,  and $\Gamma_{\pm}=\Gamma_0 e^{\pm \hbar \omega/(2k_B T)}\csch\left[\hbar \omega/(2k_B T)\right] /2$, with $\Gamma_0$ being the decay rate from the excited state to the ground state at zero temperature. The frequency $\omega$ in Eq.~(\ref{Hamiltonian}) must be redefined to include the frequency shift  caused by the presence of the thermal bath. The first term on the right hand side in Eq.~(\ref{Lindblad_eq}) describes the coherent evolution of the system, while the second term gives rise to dissipation.

By solving Eq.~(\ref{Lindblad_eq}), the density operator at an arbitrary time $\tau$ can be expressed in terms of the initial density operator $\rho_0$ in the form
\begin{equation}
	\label{channel1}
\rho_{\tau}=	\mathcal{E}(\rho_0)\equiv \sum_{j=0}^3 U E_j \rho_0 E_j^{\dagger} U^{\dagger} ,
\end{equation}
where $\{E_0,E_1,E_2,E_3\}$ are the Kraus operators for the generalized amplitude damping channel~\cite{NielsenChuang2000}. The explicit expressions for these operators are~\cite{NielsenChuang2000} 
\begin{eqnarray}
	\label{E0}
	E_0&=&\sqrt{p_+}\left(\ketbra{+}{+}+\sqrt{1-\gamma}\ketbra{-}{-}\right),\\
	\label{E1}
	E_1 &=&\sqrt{p_{+} \gamma} \tilde{\sigma}_+,\\
	\label{E2}
	E_2&=&\sqrt{1-p_+}\left(\sqrt{1-\gamma}\ketbra{+}{+}+\ketbra{-}{-}\right),\\
	\label{E3}
	E_3 &=&\sqrt{(1-p_+) \gamma} \tilde{\sigma}_-,
\end{eqnarray}
where $\smash{p_{+}=e^{- \hbar \omega/(2k_B T)}\sech\left[\hbar \omega/(2k_B T)\right] /2}$ is the thermal equilibrium probability of the excited state and $\smash{\gamma=1- e^{-\Gamma \tau}}$, with $\Gamma=\Gamma_++\Gamma_-=\Gamma_0\coth\left[\hbar \omega/(2k_B T)\right]$. Note that in the absence of dissipation, i.e., for $\Gamma_0=0$, the parameter $\gamma$ vanishes and, therefore, $\smash{E_0=\sqrt{p_+} I}$,  $\smash{E_2=\sqrt{1-p_+} I}$, and $\smash{E_1=E_3=0}$,  with $\smash{I}$ being the identity operator. Thus, Eq.~(\ref{channel1}) reduces to $\smash{\rho_{\tau}=U \rho_0 U^{\dagger}}$, which is the case considered in Ref.~\cite{Albarran_Arriagada_2020} with $\rho_0=\ketbra{\phi}{\phi}$.

Bearing in mind the above results, the protocol proposed in Ref.~\cite{Albarran_Arriagada_2020} can be adapted to the dissipative case as detailed below. The procedure involves very many iterations, so that, the state of \texttt{A} in the $k$th iteration is denoted as $\smash{\ket{\smash{\phi^{(k)}}}}$, with $k\in \mathbb{N}$. We assume that, in the first iteration, \texttt{A} is prepared in one of the basis states, 
for instance, in the state  $\smash{\ket{\phi^{(1)}}=\ket{0}}$. 
The states $\smash{\ket{0}}$ and $\smash{\ket{\phi^{(k)}}}$ are related by 
\begin{equation}
	\label{relation}
	\ket*{\phi^{(k)}}=D^{(k)}	\ket*{0}.
\end{equation}
where $\smash{D^{(k)}}$ is a unitary operator constructed inductively, starting with $\smash{D^{(1)}=I}$,  and building $\smash{D^{(k+1)}}$ out of $\smash{D^{(k)}}$ as follows:

(i) From the initial density operator $\smash{\rho_0^{(k)} =\ketbra*{\phi^{(k)}}{\phi^{(k)}} }$, we let the system evolve according to the Lindblad equation in Eq.~(\ref{Lindblad_eq}) for a time $\tau$. Call $\smash{\rho_{\tau}^{(k)}=\mathcal{E}(\rho_{0}^{(k)})} $ the density at the end of that evolution obtained by Eq.~(\ref{channel1}).

(ii) We extract information from $\smash{\rho_{\tau}^{(k)}}$ by measuring the observable $\smash{M^{(k)}=D^{(k)}\ketbra*{1}{1}D^{(k)\dagger}}$. In order to always measure the same observable $\smash{M^{(1)}=\ketbra*{1}{1}}$ for all iterations, first, we apply the unitary transformation $\smash{D^{(k)\dagger}\rho_{\tau}^{(k)}D^{(k)}}$ and, then, we measure $\smash{M^{(1)}}$. After the measurement process, the state of \texttt{A} is $\smash{\ket{m^{(k)}}}$, with $\smash{m^{(k)}}$ being the outcome of the measurement, which can be $0$ or $1$ with probabilities 
\begin{equation}
	\label{probmk0}
	P^{(k)}_0=\expval{D^{(k)\dagger}\rho_{\tau}^{(k)}D^{(k)}}{0}
\end{equation}
and $\smash{P^{(k)}_1=1-P^{(k)}_0}$, respectively. Once the measurement has been completed, the state $\smash{\ket*{\phi^{(k)}}}$ can be reconstructed from $\smash{\ket*{m^{(k)}}}$ by the unitary transformation $\smash{D^{(k)}\left[(1-m^{(k)}) I +m^{(k)}\sigma_x\right]\ket*{m^{(k)}}}$, 
with  $\smash{\sigma_x=\ketbra{0}{1}+\ketbra{1}{0}}$.

(iii) If the outcome of the measurement is $\smash{m^{(k)}=1}$, we generate three pseudo-random angles, $\smash{\alpha_x^{(k)}}$, $\smash{\alpha_y^{(k)}}$, and $\smash{\alpha_z^{(k)}}$,  uniformly distributed in the  exploration interval $\smash{\left[-w^{(k)} \pi,w^{(k)} \pi\right]}$. The width of this interval is controlled by the exploration parameter $\smash{w^{(k)} }$, which is computed inductively, starting with $\smash{w^{(1)} =1}$ (maximum width), and building $\smash{w^{(k+1)} }$ out of $\smash{w^{(k)} }$ using the rule
\begin{equation}
	\label{wk}
	w^{(k+1)}=\min\left\{1,\left[(1-m^{(k)})r  +m^{(k)} p\right]w^{(k)}\right\},
\end{equation}
with $r<1$ and $p>1$ being the reward and the punishment rates~\cite{Albarran_Arriagada_2020}, respectively.   In other words, the value of the exploration parameter is updated according to the outcome of the preceding measurement, $m^{(k)}$, and the corresponding reward or punishment. Every time the measurement outcome is $m^{(k)}=1$ a punishment is applied by increasing the value of the  exploration parameter from $w^{(k)}$ to  $w^{(k+1)}=\mathrm{min}[1,p w^{(k)}]$, thus widening the exploration interval.  The $\mathrm{min}$ function is required since the maximum value of the exploration parameter is $1$. Conversely, when the measurement outcome is $m^{(k)}=0$ a reward is granted by decreasing the value of the exploration parameter from $w^{(k)}$ to $w^{(k+1)}=r w^{(k)}$,  thus narrowing the exploration interval. Once the three pseudo-random angles $\smash{\alpha_x^{(k)}}$, $\smash{\alpha_y^{(k)}}$, and $\smash{\alpha_z^{(k)}}$ have been calculated, they are used to implement the pseudo-random rotation
\begin{equation}
	\label{random_rot}
	R^{(k)}=e^{-i \alpha_y^{(k)} \sigma_y^{(k)}/2}e^{-i \alpha_z^{(k)} \sigma_z^{(k)}/2}e^{-i \alpha_x^{(k)} \sigma_x^{(k)}/2},
\end{equation}
where the operators $\smash{\sigma_x^{(k)}}$, $\smash{\sigma_y^{(k)}}$, and $\smash{\sigma_z^{(k)}}$ are related to the Pauli operators $\sigma_x$, $\smash{\sigma_y=i (\ketbra{1}{0}-\ketbra{0}{1})}$, and $\smash{\sigma_z=\ketbra{0}{0}-\ketbra{1}{1}}$ by the unitary transformation $\smash{\sigma_{\alpha}^{(k)}=D^{(k)}\sigma_{\alpha}D^{(k)\dagger}}$, with $\alpha=x$, $y$, and $z$.

(iv)  Finally, we construct $D^{(k+1)}$  from $D^{(k)}$ as
\begin{equation}
	\label{relation1}
	D^{(k+1)}=\left[(1-m^{(k)} ) I +m^{(k)} R^{(k)}\right]D^{(k)}.
\end{equation}
 Therefore, the trade-off between exploration and exploitation, which is a characteristic of reinforcement learning~\cite{Sutton2018}, is regulated using the measurement outcome $m^{(k)}$. If the measurement outcome is $m^{(k)}=1$, the agent decides to explore and modifies its state from  $\ket*{\phi^{(k)}}$ to $\ket*{\phi^{(k+1)}}=R^{(k)}\ket*{\phi^{(k)}}$. On the contrary, if the measurement outcome is $m^{(k)}=0$, the agent decides to exploit and keeps its state invariant, i.e, $\ket*{\phi^{(k+1)}}=\ket*{\phi^{(k)}}$.

Since, according to Eq.~(\ref{random_rot}), $\smash{R^{(k)}}$ is related to the pseudo-random rotation  
\begin{equation}
	\label{random_rot_new}
	\breve{R}^{(k)}=e^{-i \alpha_y^{(k)} \sigma_y/2}e^{-i \alpha_z^{(k)} \sigma_z/2}e^{-i \alpha_x^{(k)} \sigma_x/2}
\end{equation}
by the unitary transformation $\smash{R^{(k)}=D^{(k)}\breve{R}^{(k)}D^{(k)\dagger}}$, 
Eq.~(\ref{relation1}) can also be expressed in the form
\begin{equation}
	\label{relation2}
	D^{(k+1)}=D^{(k)}\left[(1-m^{(k)} ) I +m^{(k)} \breve{R}^{(k)}\right].
\end{equation}
The recursive relations in Eqs.~(\ref{relation1}) and (\ref{relation2}) can easily be solved  to yield
\begin{equation}
	D^{(k+1)} =\prod_{j=k}^1 \left[(1-m^{(j)} ) I +m^{(j)} R^{(j)}\right]
\end{equation}
and 
\begin{equation}
	D^{(k+1)} =\prod_{j=1}^k  \left[(1-m^{(j)} ) I +m^{(j)} \breve{R}^{(j)}\right],
\end{equation}
respectively, where we use  the ordered product notation $\smash{\prod_{j=k}^1 A_j= A_k A_{k-1}\cdots A_1}$ and $\smash{\prod_{j=1}^k A_j= A_1 A_2\cdots A_k}$. 

 The dissipative protocol just described, as well as the nondissipative protocol analyzed in~\cite{Albarran_Arriagada_2020}, can be related to a Markov decision process (MDP)~\cite{MDP1,MDP2}, as in the standard reinforcement learning setup. The state space of this MDP is represented by the set of all possible quantum states, $\ket{\phi}$, in which the agent can be found. The action space is the set of all rotations, $R$,  including the identity $I$, acting on these quantum states. Since the actions or rotations act deterministically on the agent states, the MDP is deterministic~\cite{DMDP} and, consequently,  its transition model can be described in terms of a transition function rather than a transition probability.  Specifically, the transition function is $\mathcal{T}(\ket{\phi}, R)=R\ket{\phi}$, i.e., as a result of the action $R$ applied on state $\ket{\phi}$, the state changes to $\ket{\phi^{\prime}}=R\ket{\phi}$. Finally, the reward function of the MDP is described by the function $\mathcal{R}(\ket{\phi}, R)=\mathcal{R}(R)$ that takes the value $r$ if $R=I$ and $p$ if $R\neq I$. Given the functions $\mathcal{T}$ and $\mathcal{R}$, it is sufficient to know the current state $\ket{\phi}$ and the current action $R$ to determine the next state $\ket{\phi^{\prime}}$ and the corresponding reward, thus fulfilling the Markov property. Note, however, that the decision rule (iv) that determines which action to select, given the agent state at step $k$, is randomized and history-dependent~\cite{MDP1,MDP2}. Its randomness stems both from the intrinsically random nature of the measurement outcomes $\smash{m^{(k)}}$ and from the dependence of the rotations $\smash{R^{(k)}}$ on the pseudo-random angles $\smash{\alpha_x^{(k)}}$, $\smash{\alpha_y^{(k)}}$, and $\smash{\alpha_z^{(k)}}$. 	Its history dependence is due to the fact that, as described in (iii), to generate these pseudo-random angles, the value of the exploration parameter  $\smash{w^{(k)} }$ must first be determined, which implies solving the recurrence relation in Eq.~(\ref{wk}) and, thus, knowing all the previous history.

\section{Numerical results} \label{numerics}

We have carried out numerical simulations implementing the protocol presented in the previous section for a Hamiltonian of the form
\begin{equation}
	H=\frac{\hbar \omega}{2}\sigma_x,
	\label{Hamiltonian1}
\end{equation}
which corresponds to taking  $\smash{\ket{\pm}=(\ket{0}\pm\ket{1})/\sqrt{2}}$ in Eq.~(\ref{Hamiltonian}). To deal with dimensionless quantities, we have introduced the dimensionless parameters $\smash{\tilde{\tau}=\omega\tau}$,  $\smash{\tilde{\Gamma}_0=\Gamma_0/\omega}$, and $\smash{\tilde{T}=k_B T/(\hbar \omega)}$. 
To simulate numerically the measurement process appearing in item~(ii) of Sec.~\ref{protocol}, we first calculate the probability $\smash{P_0^{(k)}}$ of obtaining $\smash{m^{(k)}=0}$ using Eq.~(\ref{probmk0}). Then, we draw a pseudo-random number $\smash{\xi^{(k)}}$ uniformly distributed in the interval $\smash{[0,1]}$. If $\smash{\xi^{(k)}\leq P_0^{(k)}}$, the outcome of the measurement is $\smash{m^{(k)}=0}$, whereas if $\smash{\xi^{(k)}>P_0^{(k)}}$ the result will be $\smash{m^{(k)}=1}$.

In order to assess the accuracy of the protocol, 
we have calculated the fidelity between the state  $\smash{\ket*{\phi^{(k)}}}$ and the closest eigenvector of $H$ as a function of the number of iterations, $k$. Since it is not known \textit{a priori} whether this eigenvector is $\smash{\ket{+}}$ or $\smash{\ket{-}}$, we have considered the greater of the two values, i.e.,
\begin{equation}
	f^{(k)}=\max\left(| \braket*{+}{\phi^{(k)}}|,| \braket*{-}{\phi^{(k)}}|\right).
\end{equation}
The closer the value of $\smash{f^{(k)}}$ is to 1 as $k$ increases, the more accurate the computation of the corresponding eigenvector will be. In addition, at each iteration, the convergence of the protocol has been quantified by the exploration parameter $\smash{w^{(k)}}$ defined  in item~(iii) of Sec.~\ref{protocol}. The protocol is considered to converge if $\smash{w^{(k)}}$ approaches zero as $k$ increases. Moreover, the faster it approaches zero, the faster the convergence of the protocol. 

In the numerical simulations presented here, we have repeated the protocol a large number $N$ of realizations. 
Then, for each value of $k$, the mean fidelity
\begin{equation}
	\label{meanfidelity}
		F^{(k)}=\frac{1}{N}\sum_{j=1}^N f_j^{(k)}
\end{equation}
and the mean exploration parameter
\begin{equation}
	\label{meanexploration}
	W^{(k)}=\frac{1}{N}\sum_{j=1}^N w_j^{(k)}
\end{equation}
are obtained by averaging over the $N$ realizations, where the subscript $j$ refers to the $j$th realization of the protocol. The number of realizations considered in this paper  is $N=1000$.

\begin{figure}[h]
	\centering
	\includegraphics[width=1\linewidth]{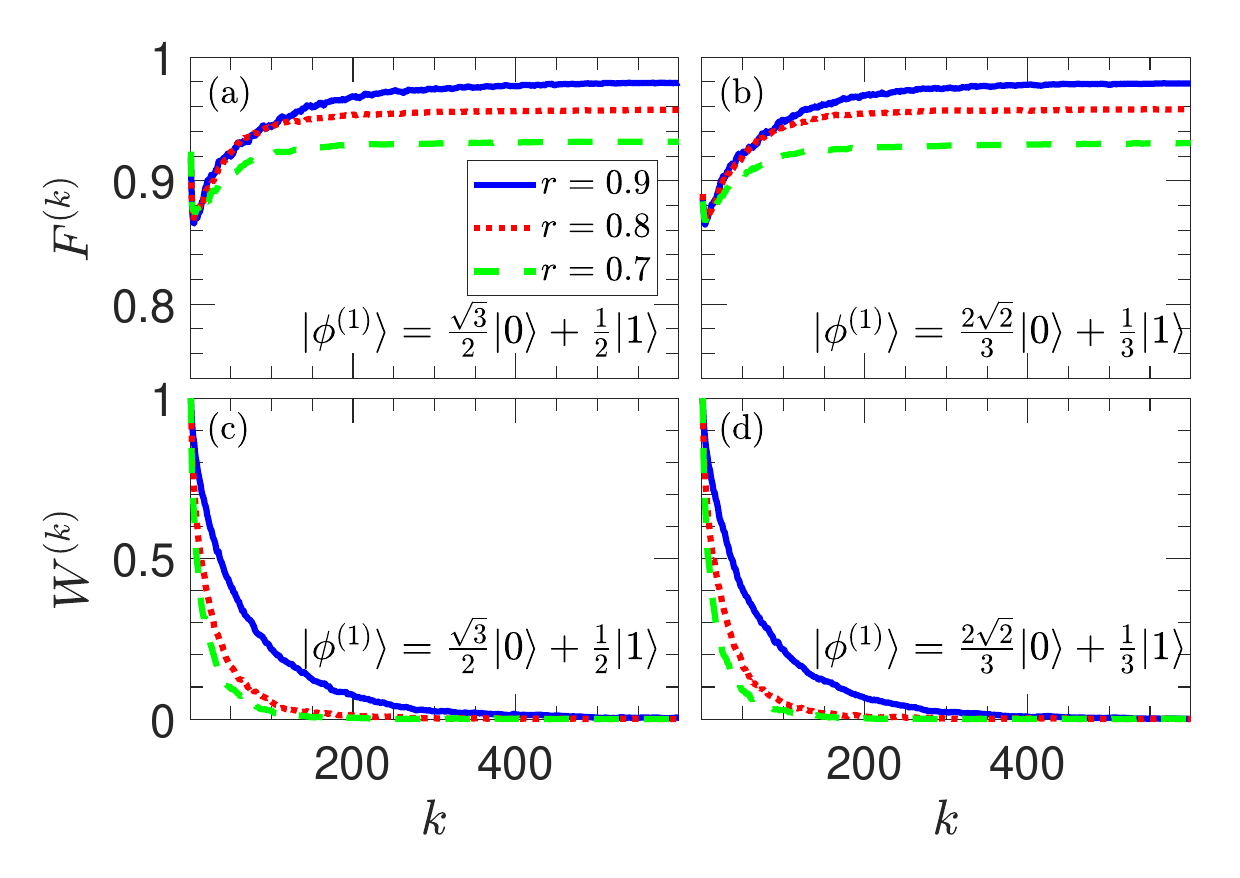}
	\caption{Dependence of the mean fidelity $F^{(k)}$ [(a) and (b)] and the mean exploration parameter $W^{(k)}$ [(c) and (d)] on the number of iterations, $k$, for different values of the reward parameter $r$. The state in the first iteration is $\smash{\ket{\smash{\phi^{(1)}}}=(\sqrt{3}\ket{0}+\ket{1}})/2$ in (a) and (c) and $\smash{\ket{\smash{\phi^{(1)}}}=(2\sqrt{2}\ket{0}+\ket{1}})/3$ in (b) and (d). The remaining parameter values are $\tilde{\Gamma}_0=0.5$, $\smash{\tilde{T}=0.3}$, $\tilde{\tau}=1$, $N=1000$, and  $p=2/r$.}
	\label{Figure1}
\end{figure}
In Fig.~\ref{Figure1} we have plotted the mean fidelity $F^{(k)}$ [(a) and (b)] and the mean exploration parameter $W^{(k)}$ [(c) and (d)] versus the number of iterations $k$. We have considered two different states for the first iteration, namely, $\smash{\ket{\smash{\phi^{(1)}}}=(\sqrt{3}\ket{0}+\ket{1}})/2$ in (a) and (c) and $\smash{\ket{\smash{\phi^{(1)}}}=(2\sqrt{2}\ket{0}+\ket{1}})/3$ in (b) and (d).	With different types of lines, we have depicted different values of the reward parameter $r$. When $r$ is  modified, the value of the punishment parameter $p$ is also varied according to the relation $p=2/r$. 
The parameters associated with thermal dissipation, i.e., the dimensionless decay rate at zero temperature and the dimensionless  temperature, are set at $\smash{\tilde{\Gamma}_0=0.5}$ and $\smash{\tilde{T}=0.3}$, respectively.  As can be seen, the values of the reward and punishment parameters affect the accuracy and convergence of the protocol. The asymptotic values of $\smash{F^{(k)}}$ in the large iteration limit are less than $0.94$ when $r=0.7$, reach values close to $0.98$ in the case of $r=0.9$, and take intermediate values for $r=0.8$. A more detailed analysis of how the nondissipative protocol is affected by varying the values of the reward and punishment parameters is provided in Ref.~\cite{Albarran_Arriagada_2020}.  Note, however, that the particular state chosen for the first iteration does not seem to play a relevant role in the accuracy and convergence of the protocol, as can be seen by comparing Figs.~\ref{Figure1}(a) and (c) with Figs.~\ref{Figure1}(b) and (d). 
 In order to focus on the effect of thermal dissipation on the performance of the protocol, we will henceforth set the reward and punishment parameters to $r=0.9$  and $p=2/r=20/9$, respectively, as well as the initial state to $\smash{\ket*{\phi^{(1)}}=\ket{0}}$. 

\begin{figure}
	\centering
	\includegraphics[width=1\linewidth]{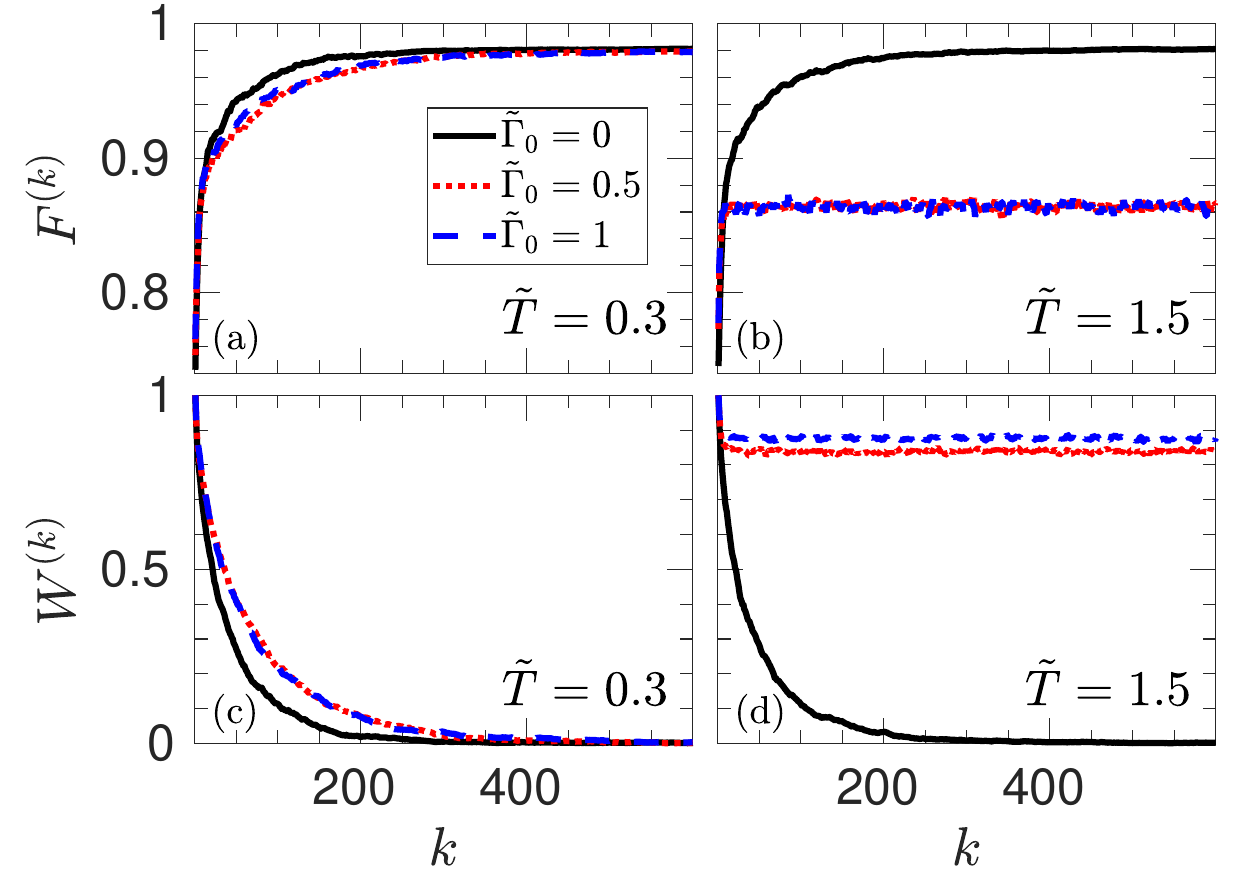}
	\caption{Dependence of the mean fidelity $F^{(k)}$ [(a) and (b)] and the mean exploration parameter $W^{(k)}$ [(c) and (d)] on the number of iterations, $k$. Black solid lines depict the nondissipative case $\tilde{\Gamma}_0=0$. The dissipative cases are indicated with red dotted lines ($\tilde{\Gamma}_0=0.5$) and blue dashed lines ($\tilde{\Gamma}_0 = 1$). The dimensionless temperature is $\smash{\tilde{T}=0.3}$  in (a) and (c) and $\smash{\tilde{T}=1.5}$ in (b) and (d). The remaining parameter values are $\tilde{\tau}=1$, $N=1000$, $r=0.9$, and  $p=20/9$.}
	\label{Figure2}
\end{figure}	

In Fig.~\ref{Figure2} we have plotted the mean fidelity $F^{(k)}$ [(a) and (b)] and the mean exploration parameter $W^{(k)}$ [(c) and (d)] versus $k$  for the parameter values indicated in the figure caption. With different types of lines, we have depicted different values of the dimensionless decay rate at zero temperature $\smash{\tilde{\Gamma}_0}$. The figure illustrates the role played by the temperature in the accuracy and convergence of the protocol. For dimensionless temperature $\smash{\tilde{T}=0.3}$ [(a) and (c)], the differences between the nondissipative case $\smash{\tilde{\Gamma}_0=0}$ (solid black line), and the dissipative cases $\smash{\tilde{\Gamma}_0=0.5}$ (dotted red line) and $\smash{\tilde{\Gamma}_0=1}$ (blue dashed line) are not very significant.  In the three cases,  the asymptotic values of $\smash{F^{(k)}}$  in the large iteration limit are very similar and quite close to $0.98$ [see Fig.~\ref{Figure2}(a)], and $W^{(k)}$ converges to zero rather quickly [see Fig.~\ref{Figure2}(c)]. Therefore, contrary to what might be expected, temperatures much lower than $\hbar \omega/k_B$ are not required for the protocol to work. Nonetheless, for sufficiently high dimensionless temperatures, such as $\smash{\tilde{T}=1.5}$, the protocol fails in the dissipative cases, as shown in Figs.~\ref{Figure2}(b) and (d). On the one hand, the asymptotic values of
$\smash{F^{(k)}}$ in the large iteration limit are substantially less than in the absence of dissipation [see Fig.~\ref{Figure2}(b)]. On the other hand,  
$\smash{W^{(k)}}$ does not approach zero as $k$ increases [see Fig.~\ref{Figure2}(d)] and, therefore, the protocol does not converge. 

\begin{figure}[h]
	\centering{\includegraphics[width=1\linewidth]{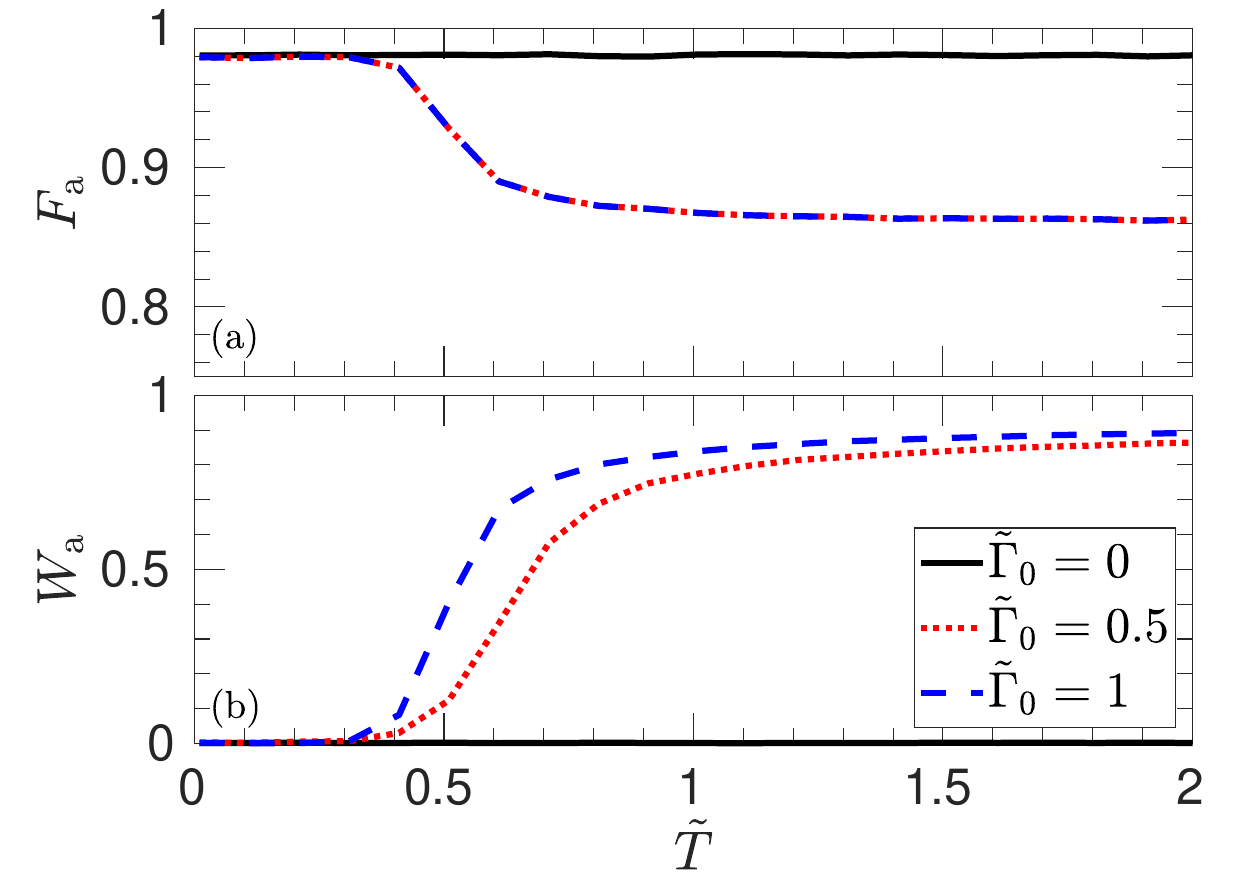}}
	\caption{Dependence of the asymptotic values  in the large iteration limit  of the mean fidelity, $F_{\mathrm{a}}$, and the mean exploration parameter, $W_{\mathrm{a}}$, on the dimensionless temperature $\tilde{T}$ for different values of $\tilde{\Gamma}_0$.   Black solid lines depict the nondissipative case $\tilde{\Gamma}_0=0$, while the dissipative cases are indicated with red dotted lines ($\tilde{\Gamma}_0=0.5$) and blue dashed lines ($\tilde{\Gamma}_0 = 1$). The remaining parameter values are $\tilde{\tau}=1$, $N=1000$, $r=0.9$, and  $p=20/9$. }
	\label{Figure3}
\end{figure}

\begin{figure}[h]
	\centering
	\includegraphics[width=1\linewidth]{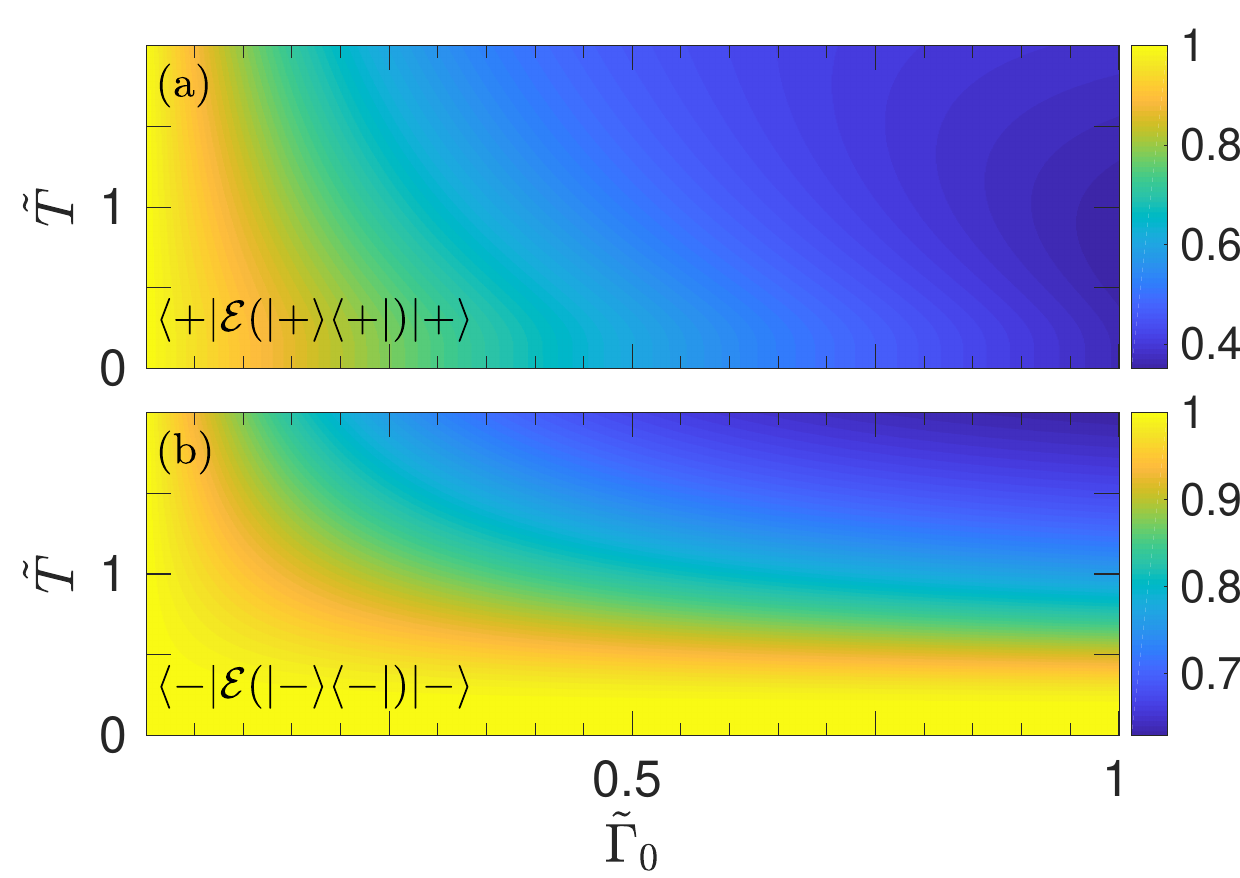}
	\caption{Dependence of the fidelities $\expval{\mathcal{E}(\ketbra{+}{+})}{+}$ in Eq.~(\ref{fidelityp})  and $\expval{\mathcal{E}(\ketbra{-}{-})}{-}$ in Eq.~(\ref{fidelityn}) on $\tilde{\Gamma}_0$ and $\tilde{T}$ for $\smash{\tilde{\tau}=1}$, $N=1000$, $r=0.9$, and  $p=20/9$.}
	\label{Figure4}
\end{figure}	

In order to determine the range of temperatures in which the protocol works well, in Fig.~\ref{Figure3} we have plotted the asymptotic values  in the large iteration limit  of the mean fidelity, $F_{\mathrm{a}}$,  and the mean exploration parameter, $W_{\mathrm{a}}$, as a function of the dimensionless temperature $\smash{\tilde{T}}$.  As can be seen, there is hardly any difference between the dissipative and nondissipative cases up to $\smash{\tilde{T}\approx 0.4}$. Above that temperature, the $F_{\mathrm{a}}$ values in the dissipative cases $\tilde{\Gamma}_0=0.5$ (red dotted line) and $\tilde{\Gamma}_0=1$ (blue dashed line) decrease abruptly and become substantially less than in the nondissipative case [see  Fig.~\ref{Figure3}(a)]. Furthermore, there is a rapid increase in $W_{\mathrm{a}}$ to nonzero values [see Fig.~\ref{Figure3}(b)], which indicates that the protocol stops converging. 

To better understand the behavior observed in Fig.~\ref{Figure3}, it is worth remembering that,  in the nondissipative case, the design of the protocol is based on the fact that the states $\ketbra{+}{+}$ and $\ketbra{-}{-}$ are fixed points of the quantum gate $\mathcal{E}$, i.e., $\mathcal{E}(\ketbra{\pm}{\pm})=\ketbra{\pm}{\pm}$.
However, in the presence of dissipation, $\ketbra{+}{+}$ and $\ketbra{-}{-}$ are no longer exact fixed points of $\mathcal{E}$,
 since the fidelities between the state $\ketbra{+}{+}$ and $\mathcal{E}(\ketbra{+}{+})$, $\expval{\mathcal{E}(\ketbra{+}{+})}{+}$, and between the state $\ketbra{-}{-}$ and $\mathcal{E}(\ketbra{-}{-})$, $\expval{\mathcal{E}(\ketbra{-}{-})}{-}$, are in general less than $1$.  Specifically, from Eq.~(\ref{channel1}), it can be shown that
\begin{align}
	\label{fidelityp}
	\expval{\mathcal{E}(\ketbra{+}{+})}{+}&= 1-(1-p_+)\gamma\\
	\label{fidelityn}
	\expval{\mathcal{E}(\ketbra{-}{-})}{-}&=1-p_+\gamma,
\end{align}
which are clearly less than 1 if $\smash{\tilde{\Gamma}_0 \neq 0}$ (dissipative case). Despite this, there may be parameter values for which the fidelities $\expval{\mathcal{E}(\ketbra{+}{+})}{+}$ and/or $\expval{\mathcal{E}(\ketbra{-}{-})}{-}$ are quite close to $1$. In that case, the corresponding states $\ketbra{+}{+}$ and/or $\ketbra{-}{-}$ would behave as approximate fixed points and the protocol would still be applicable. Figure~\ref{Figure4} shows the dependence of the fidelities $\expval{\mathcal{E}(\ketbra{+}{+})}{+}$ and $\expval{\mathcal{E}(\ketbra{-}{-})}{-}$, obtained using Eqs.~(\ref{fidelityp}) and (\ref{fidelityn}),  on  $\tilde{\Gamma}_0$ and $\tilde{T}$. As can be seen, there are regions where  these fidelities are quite close to $1$.  In particular, Fig.~\ref{Figure4}(b) shows that $\expval{\mathcal{E}(\ketbra{-}{-})}{-}$ is  quite close to $1$ up to $\smash{\tilde{T}\approx 0.4}$, which is the approximate value beyond which the protocol starts to fail in Fig.~\ref{Figure3}.

\begin{figure}[h]
	\centering
	\includegraphics[width=1\linewidth]{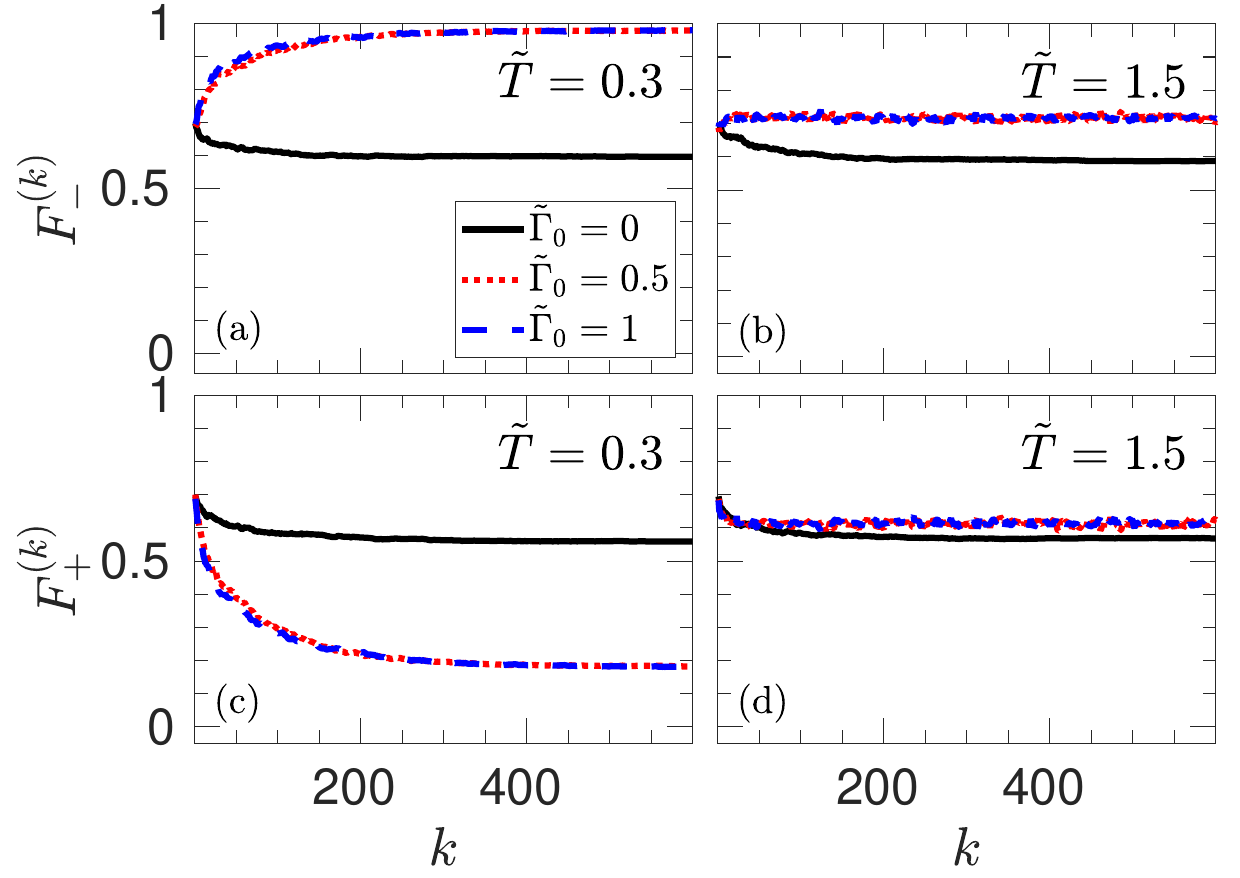}
	\caption{Dependence of the state-specific mean fidelities $\smash{F_-^{(k)}}$ [(a) and (b)] and   $\smash{F_+^{(k)}}$ [(c) and (d)] on the number of iterations, $k$. Black solid lines depict the nondissipative case $\smash{\tilde{\Gamma}_0=0}$. The dissipative cases are indicated with red dotted lines ($\smash{\tilde{\Gamma}_0=0.5}$) and blue dashed lines ($\smash{\tilde{\Gamma}_0 = 1}$).  The dimensionless temperature is $\smash{\tilde{T}=0.3}$  in (a) and (c) and $\smash{\tilde{T}=1.5}$ in (b) and (d). The remaining parameter values are $\smash{\tilde{\tau}=1}$, $N=1000$, $r=0.9$, and  $p=20/9$.}
	\label{Figure5}
\end{figure}	

From  Figs.~\ref{Figure2} and \ref{Figure3},  it may be concluded that the effect of dissipation is not very relevant for sufficiently low dimensionless temperatures. However, a more exhaustive analysis of the results reveals that it plays an important role in the protocol.  Figure~\ref{Figure5} depicts separately the mean fidelity $\smash{F^{(k)}_{-}}$ between the state $\smash{\ket*{\phi^{(k)}}}$ and the ground state $\ket{-}$ [(a) and (b)] and the mean fidelity $\smash{F^{(k)}_{+}}$ between the state $\smash{\ket*{\phi^{(k)}}}$ and the excited state $\ket{+}$ [(c) and (d)], defined as
\begin{equation}
	\label{meanstatefidelity}
	F^{(k)}_{\mp}=\frac{1}{N}\sum_{j=1}^N |\braket*{\mp}{\phi_j^{(k)}}|.
\end{equation}
While, for $\smash{\tilde{T}=0.3}$, the differences between the dissipative and nondissipative cases are not too significant when comparing the mean fidelity $F^{(k)}$ [see Fig.~\ref{Figure2}(a)], they become rather noticeable when comparing the state-specific mean fidelities $\smash{F^{(k)}_{-}}$ and $\smash{F^{(k)}_{+}}$  [see Figs.~\ref{Figure5}(a) and (c)]. In the nondissipative case, $\smash{F^{(k)}_{-}}$ and $\smash{F^{(k)}_{+}}$ are almost indistinguishable [black solid lines in Figs.~\ref{Figure5}(a) and (c)], as is to be expected from the fact that $\ketbra{+}{+}$ and $\ketbra{-}{-}$ are equivalent fixed points of $\mathcal{E}$ in the absence of dissipation. By contrast, the presence of dissipation substantially affects the state-specific fidelities, increasing $\smash{F^{(k)}_{-}}$ and decreasing $\smash{F^{(k)}_{+}}$ [red dotted lines and blue dashed lines in Figs.~\ref{Figure5}(a) and (c)]. This is because, for $\smash{\tilde{T}=0.3}$, $\smash{\tilde{\Gamma}_0=0.5}$ and $\smash{\tilde{\Gamma}_0=1}$, $\ketbra{-}{-}$ is an approximate fixed point of $\mathcal{E}$ but $\ketbra{+}{+}$ is not [see Fig.~\ref{Figure4}].
As the temperature increases, this difference becomes less appreciable, as can be seen in Figs.~\ref{Figure5}(b) and (d).

\begin{figure}[h]
	\centering
	\includegraphics[width=1\linewidth]{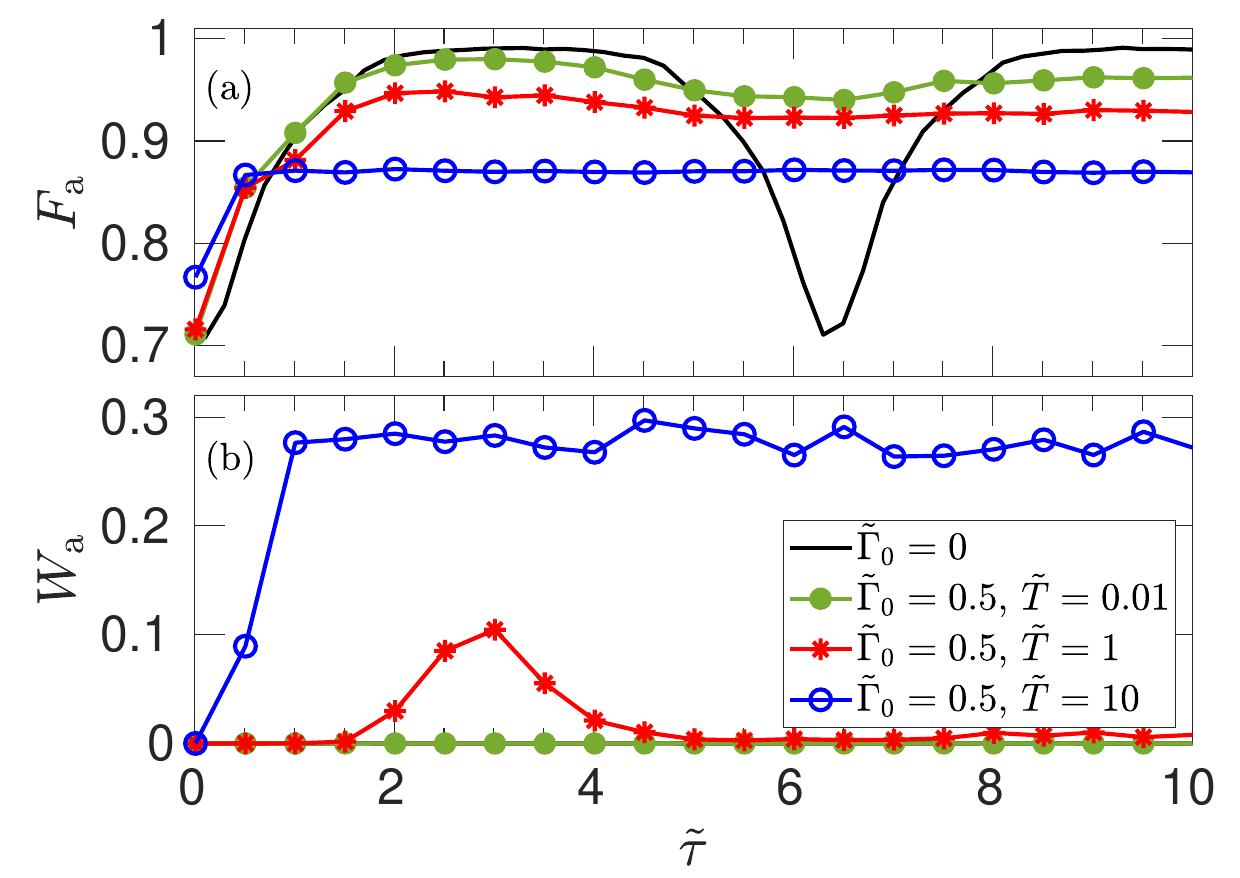}
	\caption{Dependence of the asymptotic values  in the large iteration limit  of the mean fidelity, $F_{\mathrm{a}}$, and the mean exploration parameter, $W_{\mathrm{a}}$, on the dimensionless evolution time $\tilde{\tau}$ for different values of $\tilde{\Gamma}_0$ and $\tilde{T}$.   Black solid lines depict the nondissipative case $\smash{\tilde{\Gamma}_0=0}$. In (b) the black solid line is indistinguishable from the abscissa-axis. The dissipative cases are indicated with green filled circles ($\smash{\tilde{\Gamma}_0=0.5}$ and $\tilde{T}=0.01$), red stars ($\smash{\tilde{\Gamma}_0=0.5}$ and $\tilde{T}=1$), and blue open circles ($\smash{\tilde{\Gamma}_0 = 0.5}$ and $\tilde{T}=10$).   The remaining parameter values are $N=1000$, $r=0.9$, and  $p=20/9$.}
	\label{Figure6}
\end{figure}	

One important aspect, not considered so far, is the dependence of the protocol results on the evolution time $\tau$. As mentioned in Sec.~\ref{protocol}, in the absence of dissipation, i.e., for $\Gamma_0=0$, the action of \texttt{E} on \texttt{A} is represented by the quantum channel $\smash{\mathcal{E}(\rho_0)= e^{-i \tau H/\hbar}\rho_0  e^{i \tau H/\hbar}}$. For the case of the Hamiltonian $H$ in Eq.~(\ref{Hamiltonian1}), this quantum channel has the explicit form 
\begin{equation}
	\label{NDE}
\begin{split}
	\mathcal{E}(\rho_0)&=\frac{1}{2}(\rho_0+\sigma_x\rho_0\sigma_x) \\ 
	&\quad +\frac{\cos(\tilde{\tau})}{2}(\rho_0-\sigma_x\rho_0\sigma_x) +\frac{i \sin(\tilde{\tau})}{2}[\rho_0,\sigma_x],
\end{split}
\end{equation}
which is a periodic function of the dimensionless evolution time $\smash{\tilde{\tau}}$ with period $2\pi$.
This periodicity is clearly visible in Fig.~\ref{Figure6}(a), where the dependence of the asymptotic values in the large iteration limit  of the mean fidelity, $F_{\mathrm{a}}$, on $\smash{\tilde{\tau}}$ in the nondissipative case is depicted with a solid line. However, the periodicity is not evident in the behavior of $W_{\mathrm{a}}$, which always remains close to $0$ and indistinguishable from the abscissa-axis, as shown in Fig.~\ref{Figure6}(b). Of particular interest is the case in which $\smash{\tilde{\tau}}$ is an integer multiple of $2\pi$. In this case,  it follows from Eq.~(\ref{NDE}) that \texttt{E} does not modify the state of \texttt{A} and, therefore, no learning is possible. Thus, for this case the mean fidelity $\smash{F^{(k)}}$ is independent of the number of iterations, $k$ and, consequently,  $\smash{F_{\mathrm{a}}=F^{(1)}=1/\sqrt{2}\approx 0.71}$, as it can be observed in Fig.~\ref{Figure6}(a) (black solid line). 

In the presence of dissipation,  i.e., for $\Gamma_0\neq0$, the quantum channel $\smash{\mathcal{E}(\rho_0)}$ in Eq.~(\ref{channel1}) ceases to be periodic in $\smash{\tilde{\tau}}$ due to the exponential dependence of the parameter $\gamma$ on the evolution time $\tau$ (see Sec.~\ref{protocol}). As a consequence, the oscillations observed in Fig.~\ref{Figure6}(a) for the nondissipative case (black solid line) are notably attenuated and, beyond a certain value of $\smash{\tilde{\tau}}$, $F_{\mathrm{a}}$ becomes almost independent of $\smash{\tilde{\tau}}$ [see green filled circles, red stars, and blue open circles in Fig.~\ref{Figure6}(a)]. The effect of temperature on the behavior of $F_{\mathrm{a}}$  is quite revealing. The long $\smash{\tilde{\tau}}$ values of $F_{\mathrm{a}}$ for the dissipative case decrease as the temperature increases, as expected. Nonetheless, the $F_{\mathrm{a}}$ values are substantially larger than those for the nondissipative case for  $\smash{\tilde{\tau}}$  values around the integer multiples of $2\pi$. Therefore, under these circumstances, dissipation plays a positive role in the protocol fidelity. Note that for not too high temperatures, the convergence of the protocol gauged by $W_{\mathrm{a}}$ is still guaranteed around the above mentioned $\smash{\tilde{\tau}}$ values [see green filled circles and red stars in Fig.~\ref{Figure6}(b)].

\section{Conclusions} \label{Conclusions}

 In this study, we investigated the impact of thermal dissipation on a protocol for quantum reinforcement learning. To achieve this, we introduced a Lindblad dynamics for the density operator of a two-state system in contact with a thermal bath. First, we conducted a comprehensive theoretical analysis of the protocol and established the relevant quantities that characterize it. Subsequently, we performed a numerical analysis to demonstrate the influence of thermal dissipation on the protocol's performance. It is important to understand how thermal dissipation affects quantum reinforcement learning because machine learning algorithms are typically not implemented in machines that operate at absolute zero temperature, and therefore, thermal effects cannot be avoided in real-world situations.
 
 The fundamental difference between the methodology presented in this study and that considered in Ref.~\cite{Albarran_Arriagada_2020} lies in the way in which the agent interacts with the environment to gather information from it. In Ref.~\cite{Albarran_Arriagada_2020}, such interaction is free from dissipation and is therefore described by the unitary transformation $\smash{\rho_{\tau}=U \rho_0 U^{\dagger}}$. However, in this study, the interaction involves a certain degree of thermal dissipation which causes it to deviate from unitarity  [see Eq.~(\ref{channel1})]. This makes the description more realistic in our case as experiments always take place at nonzero temperatures and, therefore, in the presence of some degree of thermal dissipation.

The main conclusions of this work are: 

(i) For sufficiently low temperatures, dissipation does not necessarily have a negative effect on the accuracy of the protocol gauged by the mean fidelity $\smash{F^{(k)}}$. In fact, our analysis shows that there are ranges of parameter values for which the dissipative protocol performs better than the nondissipative one, as can be seen in Fig.~\ref{Figure6}. These results might be of interest for the experimental implementation of this type protocols, since sometimes the presence of dissipation is unavoidable in an actual experiment.

(ii) Dissipation is particularly relevant when evaluating the state-specific mean fidelities $\smash{F^{(k)}_{-}}$ and $\smash{F^{(k)}_{+}}$. The influence of dissipation on these fidelities depends dramatically on which state-specific is considered. While in the nondissipative case $\smash{F^{(k)}_{-}}$ and $\smash{F^{(k)}_{+}}$ are almost indistinguishable, in the dissipative case they can be quite different. Specifically, for sufficiently low temperatures,  the mean fidelity with respect to the ground state, $\smash{F^{(k)}_{-}}$, is much higher than the one obtained in the absence of dissipation, whereas the mean fidelity with respect to the excited state, $\smash{F^{(k)}_{+}}$, is much lower.  Therefore, if in an actual experiment one were interested in specifically computing the ground state, the presence of dissipation might be useful.

\section*{Acknowledgements}

We acknowledge funding by the Junta de Andaluc\'ia and FEDER (P20-00617 and US-1380840) and by the Spanish Ministry of Science, Innovation, and Universities under grant Nos. PID2019-104002GB-C21 and PID2019-104002GB-C22.



%

\end{document}